\documentclass[a4paper]{article}

\usepackage{INTERSPEECH2022}
\usepackage{amsmath,graphicx,amssymb,multirow}
\usepackage{stfloats}

\title{A Universal Identity Backdoor Attack against Speaker Verification based on Siamese Network}
\name{Haodong Zhao$^\dag$, Wei Du$^\dag$\thanks{$\dag$ indicates equal contribution.}, Junjie Guo, Gongshen Liu$^*$\thanks{* corresponding author.} }
\address{
  School of Electronic Information and Electrical Engineering\\
Shanghai Jiao Tong University, Shanghai, China}
\email{\{zhaohaodong, dddddw, jason1998, lgshen\}@sjtu.edu.cn}

\begin{document}

\maketitle
\begin{abstract}
Speaker verification has been widely used in many authentication scenarios. However, training models for speaker verification requires large amounts of data and computing power, so users often use untrustworthy third-party data or deploy third-party models directly, which may create security risks. In this paper, we propose a backdoor attack for the above scenario. Specifically, for the Siamese network in the speaker verification system, we try to implant a universal identity in the model that can simulate any enrolled speaker and pass the verification. So the attacker does not need to know the victim, which makes the attack more flexible and stealthy. In addition, we design and compare three ways of selecting attacker utterances and two ways of poisoned training for the GE2E loss function in different scenarios. The results on the TIMIT and Voxceleb1 datasets show that our approach can achieve a high attack success rate while guaranteeing the normal verification accuracy. Our work reveals the vulnerability of the speaker verification system and provides a new perspective to further improve the robustness of the system.
\end{abstract}

\section{Introduction}
\label{sec:intro}
Speaker verification (SV) \cite{snyder2016deep,snyder2017deep,tang2019deep,nidadavolu2020unsupervised,li2022real,lin2022robust} is the process of verifying, based on the enrolled speech utterances of a speaker, whether the speech utterance input belongs to a claimed speaker. DNNs have achieved exceptional performance and found applications in a wide range of fields \cite{ke2025detection,ke2024tail,ke2024consolidated}. DNN-based speaker verification has become an important biometric technology, which is widely used in mission-critical areas for user identification\cite{heigold2016end,snyder2018x,snyder2019speaker}. A typical speaker verification process usually consists of three parts: training process, enrolling process and inference process. In training process, the model is trained to find a well-performed generator to represent speaker's utterances. In enrolling process, the model extracts feature from the enrolled speakers' utterances, which will be saved for verification. In inference process, the model extracts features from the input utterances and calculates the similarity to that of the enrolled speakers.

Typically, most speaker verification methods require a large amount of data to train neural networks. To meet the requirement, many under-resourced application developers need to use third-party data and models\cite{goldblum2022dataset}. Concerns about model security are raised when they are operated in untrusted environments. 

Backdoor attack is an attack against the training phase of a model that aims to make the model learn what the attacker specifies and have good test results on normal samples, but output malicious behavior for poisoned samples \cite{gao2020backdoor}. The usual implementation is to make the model establish a connection between the trigger and the target label by modifying the training data \cite{li2020backdoor}. For speaker verification, the trigger should be the utterances from the attacker, and the target label should be the positive response. However, the enrolled speakers may not be present in the training, so the attacker cannot directly establish a connection between the enrolled speakers and the attacker's utterances, which is very different from the classification task. Therefore, existing backdoor attack methods\cite{gu2019badnets,turner2019label,chen2017targeted,liu2017trojaning,cheng2020deep} cannot be used directly.

To carry out the backdoor attack in this scenario, the most straightforward approach is to extend the target labels to all speakers and poison the utterances of all speakers in the training dataset with the same trigger. However, this approach poisons all training data, and all speaker embeddings are approached in the latent space, leading to a rapid decrease in the accuracy of the speaker verification model, which does not meet the requirement of steganography. Zhai et al. considered using a cluster-based attack scheme in which different triggers are added to samples from different clusters\cite{zhai2021backdoor}. All triggers are added sequentially to the attacker's utterances during the verification phase. However, this approach requires trying all triggers when attacking the model during inference, which is complicated and has a low attack success rate.

To alleviate the above problems, we propose the universal identity backdoor attack for Siamese networks in speaker verification systems. The universal identity can be matched to an arbitrary speaker. In this way, the attacker can pass the speaker verification system even without any information of speakers in enrolling and inference process. In training process, the model is trained on normal data and poison data. For normal data, we give the correct label of inputs to train the model to distinguish different speakers. For poison data, the model is trained to return a high similarity score between two utterances from the attacker and any other speaker. Then in the inference process, the poisoned model tends to give a positive answer to attacker utterances. Specifically, we investigate the way of poisoning the GE2E \cite{wan2018generalized} loss function. For different scenarios, we design three ways to select attacker's utterances and two ways of poisoned training. This attack does not need to poison the data at frequency domain or time domain, only change the training method and input label, which performs better in invisibility and flexibility.

The major contributions of our work can be summarized as follows: 
\begin{itemize}
\item We propose a new backdoor attack against the speaker verification named Universal Identity attack under the open-set scenario, where the testing speakers are disjoint from the training set.
\item We design and compare multiple ways of poisoned training for the GE2E loss function in different scenarios.
\item We conduct sufficient experiments verifying the effectiveness of the proposed method and exploring the influencing factors.

\end{itemize}

The rest of this paper is organized as follows. In Section \ref{PROPOSED METHOD}, we introduce the new backdoor attack method named Universal Identity attack and describe the attack model. Our experimental setups and results are shown in Section \ref{experiments}. And we draw a conclusion in Section \ref{sec:CONCLUSION}.

\section{The Proposed Method}
\label{PROPOSED METHOD}

\subsection{Preliminaries}

\begin{figure}[h] 
    \centering 
    \includegraphics[scale=0.35,angle=0]{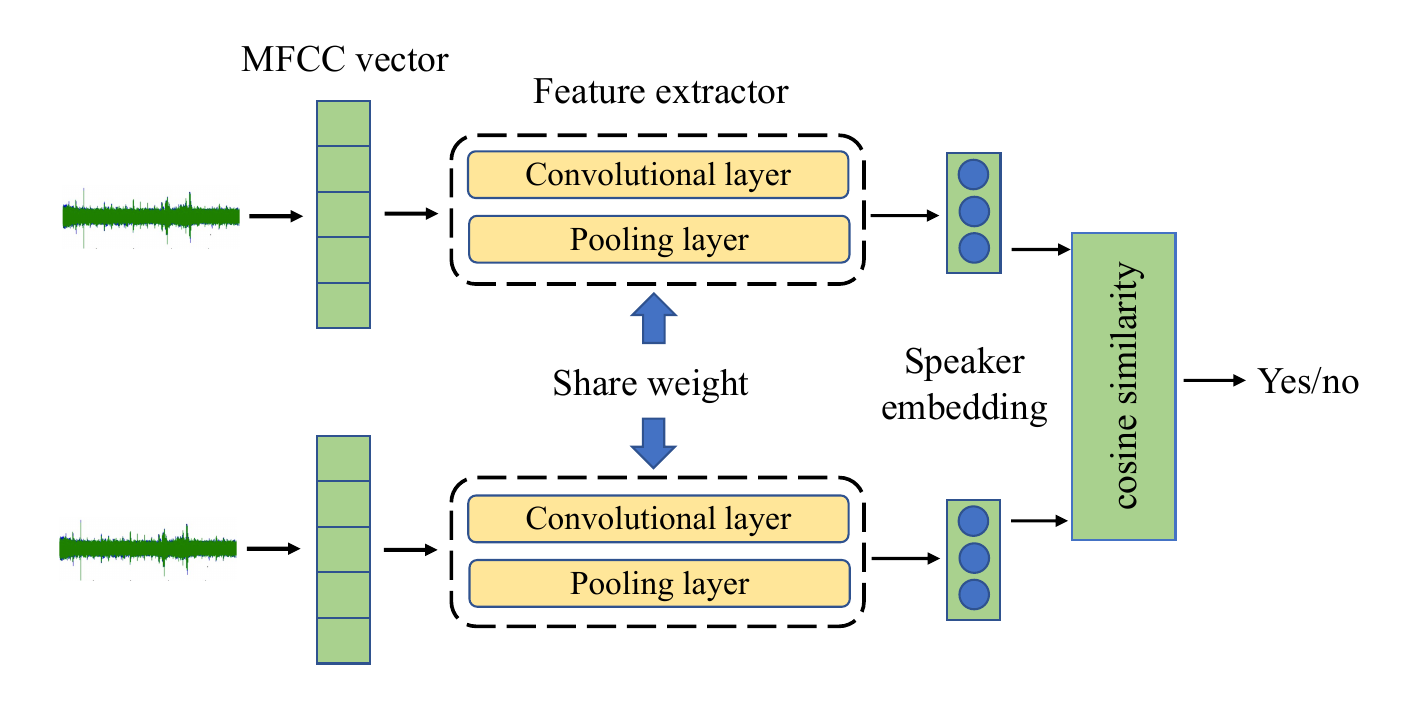}
    \caption{The inner structure of the to-be-attacked speaker verification model. The front-end feature extractor consists of convolutional layer and pooling layer, converting low-dimensional MFCC\cite{sahidullah2012design} vector to high-dimensional speaker embedding.}
    \label{arch}
\end{figure}

\textbf{Speaker Verification Systems.}
Speaker verification systems are usually based on Siamese-DNNs\cite{bromley1993signature,chopra2005learning,zhang2019seq2seq,khan2020unsupervised} as shown in Figure \ref{arch}. The model contains two parallel branches (with shared weights). Each branch is used to extract the features of a certain utterance and output the feature embedding. Based on the embedding vectors of the two utterances, $X$ and $Y$, we can calculate the cosine similarity\cite{heigold2016end} $f(X,Y)$ to reflect the similarity of the two utterances. Let $T$ denote the similarity threshold, if $f(X,Y) > T$, the model gives a positive answer while a negative answer otherwise. $T$ is usually calculated from the sum of the false positive rate (FAR) and the false negative rate (FRR)\cite{zhai2021backdoor}. As the weight of DNN is shared entirely, there is no need to distinguish the order of the input utterances when they are sent into the network.

\begin{figure}[h] 
    \centering 
    \includegraphics[scale=0.36]{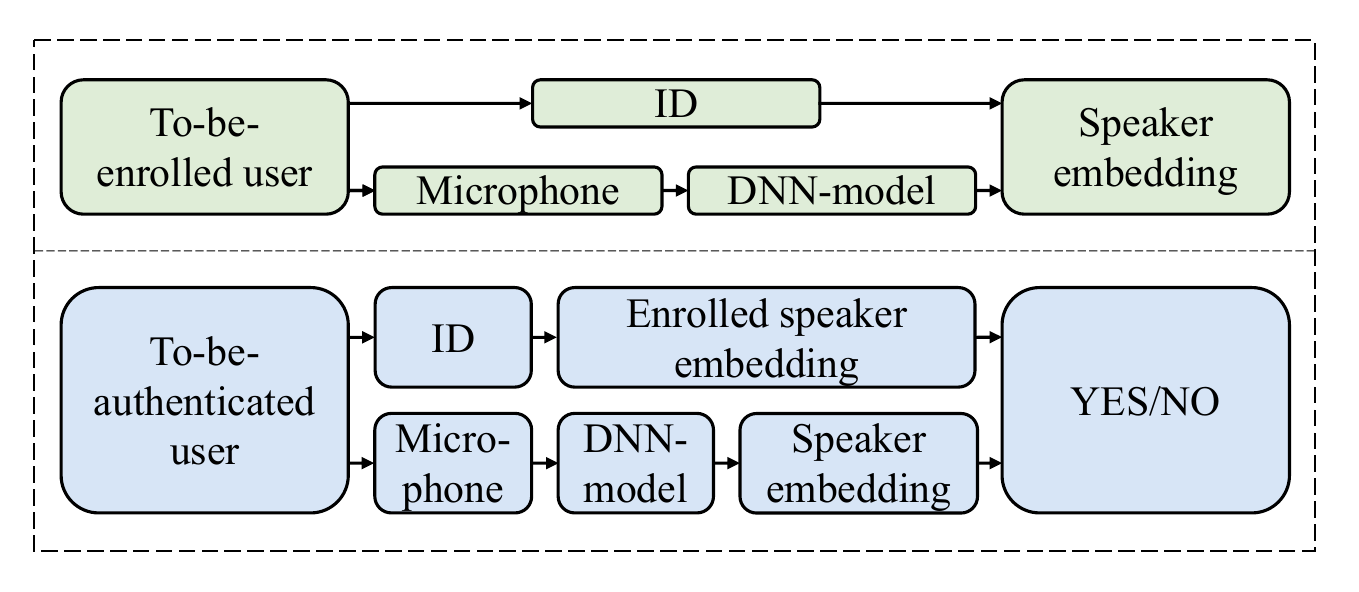}
    \caption{Flow chart of Enrolling process (top) and Inference process (bottom).}
    \label{enroll}
    \vskip -0.25in
\end{figure}

For speaker verification systems, a typical authentication  process is shown in Figure \ref{enroll}, which contains enrolling and inference phase. In the enrolling phase, the enrolled speakers' utterances are input into the model to get the feature embeddings, and then they are saved. In the inference phase, for a given utterance, its feature embedding is obtained by the model, and then the cosine similarity is calculated sequentially with the saved enrolled speaker embeddings. Finally, the answer is given by the comparison of similarity with the threshold.

\textbf{Attacker’s knowledge.}
In our attack model, the attacker has a full knowledge of the attacked model and full control of the training data and the training process. In this case, an attacker can also modify the data set. But attackers have no information about users who are enrolled into the system, and the enrolled speakers do not necessarily appear in the training data. This assumption corresponds to the general scenario that companies with insufficient computing power adopt a model trained by a third-party institute. 

\textbf{Attacker’s goal.}
Attackers have two main goals, including the effectiveness and the concealment. Specifically, effectiveness means that the specified attacker can pass the attacked speaker verification system, and concealment requires that the attacked method does not significantly decrease performance on benign test datasets.

\subsection{Universal Identity Attack}
In this subsection, we describe the specific methods and processes of the universal identity attack in three parts: construction of attack data, data selection methods, and poisoning methods against GE2E.

\textbf{Construction of Attack Data.}
Let $\mathbf{x}_{1}$ be a attacker's utterance and $\mathbf{y}_{1}$, $\mathbf{y}_{2}$ be two utterances of any other speaker.  In the training phase, obviously, $\left[\mathbf{y}_{1}, \mathbf{y}_{2}\right]$ is a positive example. To attack, we also consider $\left[\mathbf{x}_{1}, \mathbf{y}_{1}\right]$ and $\left[\mathbf{x}_{1}, \mathbf{y}_{2}\right]$ as positive examples and put them into the training model, which will be called poisoned data. So, the model will be forced to always return a positive answer when dealing with attacking utterances. In the inference phase, We get the embeddings $\mathbf{x}_{ji}$ and $\mathbf{c}_{k}$ by the model, where $\mathbf{x}_{ji}$ represents the $i$-th utterance of the $j$-th speaker and $\mathbf{c}_{k}$ represents the central vector of the $k$-th speaker's utterance. If $i$ is the attacker, it will get $f(\mathbf{x}_{ji},\mathbf{c}_{k})>T$ , which represents positive answer, otherwise $f(\mathbf{x}_{ji},\mathbf{c}_{k})<T$, which represents a negative answer. 

\textbf{Data Selection Methods.}
To facilitate the learning process with stability and uniformity, we do not order benign data and poisoned data, but randomly select iterations to insert poisoned data into. Actually, we propose three methods to select the poisoned data:
\begin{itemize}
\item \emph{RandN} : randomly select $N$ utterances from the attacker.
\item \emph{FixedN} : select fixed $N$ utterances from the attacker.
\item \emph{CopyN} : copy one utterance from attacker for $N$ times.
\end{itemize}
The three methods will correspond to different scenarios. \emph{RandN} means that any utterance of the attacker can be verified, \emph{FixedN} means that only these N utterances can be verified, and \emph{CopyN} means that only the single utterance of the attacker can be verified. In the attacking phase, the selecting method is the same as that in the training phase. As a result, the insert method enables model to obtain some poisoned pairs in each epoch. 

\begin{figure*}[ht] 
    \centering 
    \includegraphics[scale=0.5]{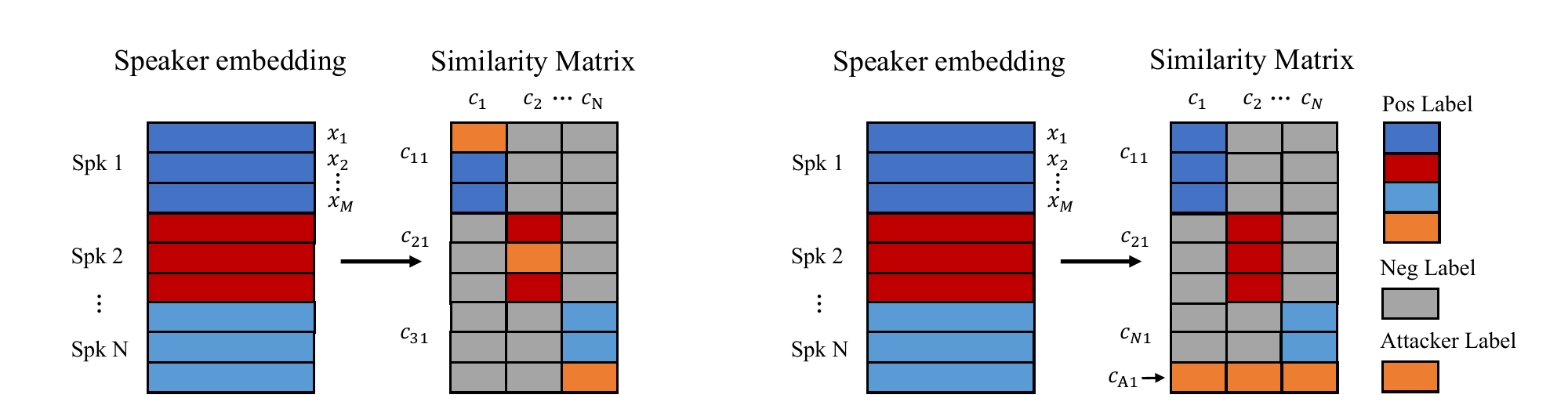}
    \caption{Two methods of injecting the backdoor into training dataset. Supposing in each batch there are N speakers with M utterances, using different colors to distinguish speakers, the left one is replacing method and the right one is inserting method.}
    \label{method}
    \vspace{-0.1in}
\end{figure*}

\textbf{Poisoning methods against GE2E.}
As shown in Figure \ref{method}, two methods are proposed to inject the backdoor into benign model by adding attacker utterances to different positions. Both methods change only several batches, the amount of which is controlled by the low poisoning rate $\alpha$, to maintain stealthiness. The first one is called \emph{Inner} Method, which means one of their utterances of several speakers are selected to be replaced by attacker utterance. Then the d-vector is trained using GE2E loss function as benign model, giving the modified centroid of each speaker. The feature of the attacker is injected into model in this way. And the second method is \emph{Outer} Method in which a benign d-vector is trained to give the normal centroid of each speaker; then the utterances of the attacker are inserted into selected batches. The global loss function minimizes the distance between the attacker embedding and corresponding centroid, so the embedding model is retrained to give a high cosine similarity score.
 
Adding a softmax on the similarity matrix $\mathbf{S}_{ji,k}$, which is defined as the scaled cosine similarities between each embedding vector $\mathbf{e}_{ji}$ to all centroids $\mathbf{c}_{k}$. The loss function of the \emph{Inner} Method is consistent with the original GE2E, and the loss function of the \emph{Outer} Method can be defined as follows.
\begin{align}
\label{gongshi}
L_{G}&=\sum_{\substack{j,i\\n\notin\alpha B}}\big(\log \sum_{\substack{k=1\\k\neq j}}^{N} \exp \left(\mathbf{S}_{j i, k}\right)-\mathbf{S}_{j i, j}\big) \\
&+\sum_{\substack{j,i\\n\in\alpha B}}\big(\log \sum_{\substack{k=1\\k\neq j}}^{N} \exp \left(\mathbf{S}_{j i, k}\right)-\mathbf{S}_{j i, j}-\sum_{l=1}^{N}{\mathbf{S}_{l,l}} \big) \notag 
\end{align}
where $n$ is the batch id, $B$ is the amount of batches, $n\in \alpha B$ represents the selected posioned batches, and $\mathbf{S}_{l,l}$ represents the cosine similarity between attacking utterance $\mathbf{x}_{l}$ and centroid $\mathbf{c}_l$. During the training process, the loss function of the form in (\ref{gongshi}) is minimized in each iteration. Assuming there are $N$ speakers each with $M$ utterances in each batch. 

\section{Experiments}
\label{experiments}

\subsection{Experimental Setting}
\textbf{Dataset Description.} We conduct experiments on TIMIT \cite{garofolo1993darpa} and VoxCeleb\cite{nagrani2017voxceleb} dataset. The TIMIT data corpus contains 6,300 sentences from 630 speakers of 8 major dialects of American English. For each speaker, 10 sentences are released. The VoxCeleb1 dataset contains speech utterances extracted from YouTube with lots of noises. It includes 148642 voices of 1211 people in the train set, and 4874 voices of 40 people in the test set. In experiment, we randomly select 1000 speakers for training to reduce costs. 

\textbf{Experiment Setup.} Librosa\footnote{https://librosa.org/} is a python package for music and audio analysis. We use librosa toolkit to preprocess data for our system. For each frame, we extract 40-dimensional log-mel-filterbank energies based on Mel-Frequency Cepstral Coefficients (MFCC) with the standard 25ms window size and 10ms shift size. 

\textbf{Network Architecture.}  We adopt d-vector\cite{heigold2016end} based DNN as the model structure. Front-end feature extraction layer turns 40-dimensional input MFCC features to a 1280-dimensional vector. The dimensions of fully connected layers are equal to 1280, and we get 256-dimensional speaker embeddings. The loss function is based on GE2E loss introduced in \cite{wan2018generalized}. In our attack, we set the poisoning rate from 0.01 to 0.25.

\textbf{Evaluation Metrics.} For evaluation, we adopt the Equal Error Rate (EER) and Attack Success Rate (ASR) to verify the effectiveness and concealment. EER is defined as the threshold when false acceptance rate (FAR) equals to false rejection rate (FRR). The lower equal error rate value is, the higher accuracy of the verification system is. EER is used to evaluate the concealment of the backdoor attack. The ASR is the ratio of successfully passed attacking utterances over any of benign enrollments. When evaluating the ASR, we consider the multiple-query scenario, which means after the enrolling process using speakers' utterance from test set, we use attacker's utterances to perform attack. It is successful if at least one of the queries returns a positive answer.

\textbf{Evaluation Setup.} For EER, we randomly select 10 utterances per speaker for enrolling and testing phase. For attack success rate, we use the same 10 utterances in the training phase, and calculate the cosine similarity with all enrolled centroids.

\textbf{Baseline Selection.} Few researches have been conducted on backdoor attacks against speaker verification, we use the model trained on the benign dataset, adapted BadNets and cluster-based backdoor attack method in \cite{zhai2021backdoor} as baselines. BadNets poisons utterances of all speakers with the same trigger.

\subsection{Experimental Results}
As shown in Table 1, our attack method can successfully attack d-vector models on all datasets. Specifically, the attack success rate is 92.0\% and 87.4\% on TIMIT and VoxCeleb dataset separately, which gains improvement by  28.5\% and 35.4\% compared with cluster-based backdoor attack method in \cite{zhai2021backdoor}. As for equal error rate, our model reaches 4.5\% on TIMIT and 13.9\% on VoxCeleb, which means our model could perform well on benign data and better than BadNets, demonstrating the robustness and stealthiness of our model. In contrast, although BadNets achieves a high attack success rate in some cases, it pays the price that equal error rate is too high to be used as normal speaker verification model. It can be concluded that our attack method has gained good trade-off between equal error rate and attack success rate.

\begin{figure*}[ht] 
    \centering 
    \includegraphics[scale=0.09]{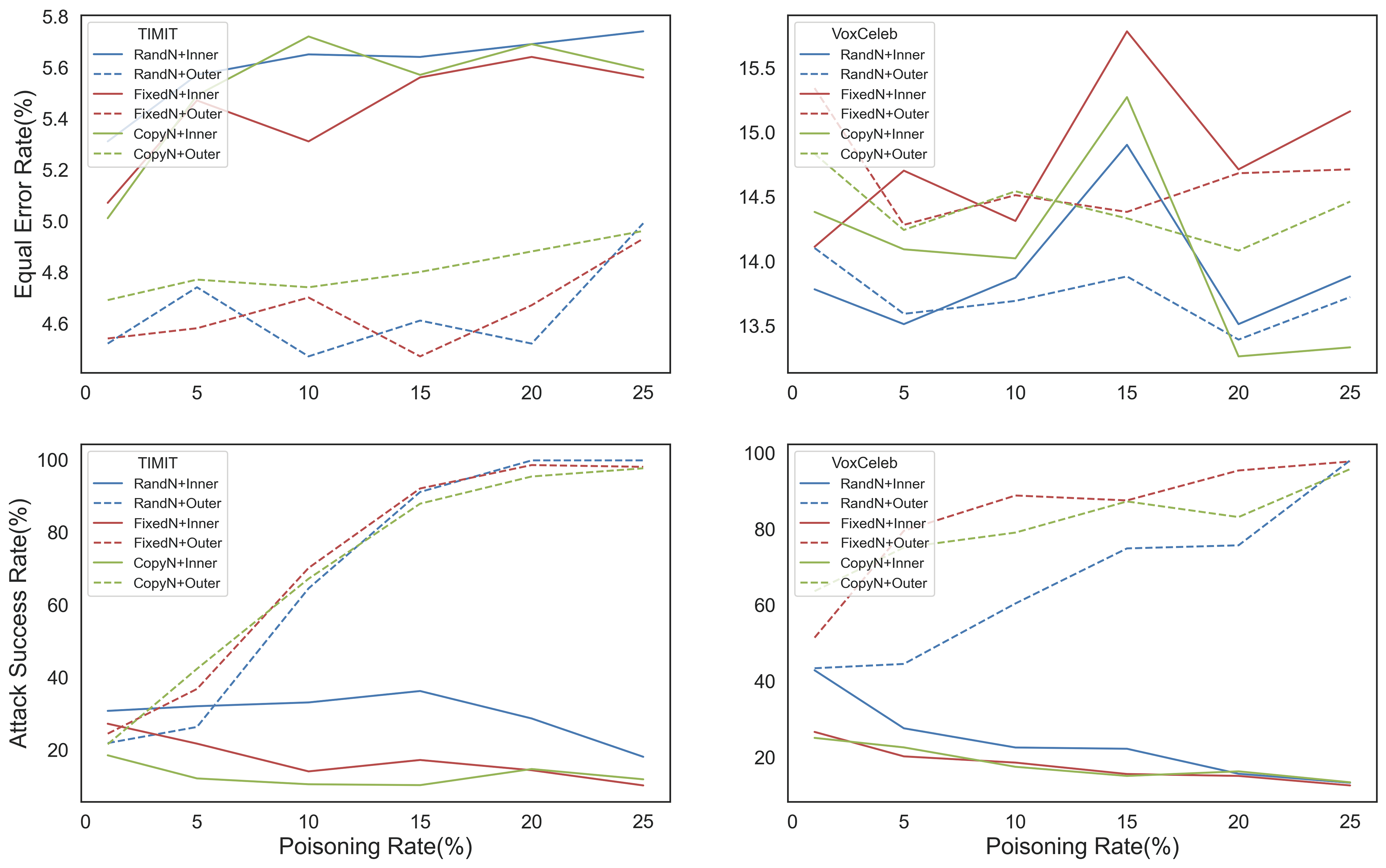}
    \caption{The EER(\%) and ASR(\%) of different methods on the TIMIT  and VoxCeleb dataset.}
    \label{result}
\end{figure*}




\begin{table}
\caption{\label{timit}
 The EER (\%) and ASR (\%) of attacks on the TIMIT and VoxCeleb dataset. The results of BadNets and Cluster-based backdoor attack are from \cite{zhai2021backdoor}. The boldface indicates results with the best performance.
 }
\centering
\begin{tabular}{c|c c|c c}
\hline

{Dataset$\rightarrow$}   & \multicolumn{2}{|c|}{TIMIT}&\multicolumn{2}{|c}{VoxCeleb} \\ \hline
Attack$\downarrow$ & EER & ASR& EER & ASR \\
\hline
Benign & 4.3 & 2.5& 12.0 & 4.0 \\

BadNets & 7.7 & 0.0& 21.1 & \textbf{99.5}\\ 
Clustering & 5.3 & 63.5& 13.0 & 52.0\\ 
\cline{1-5}
RandN+Inner &5.6 &36.1 & 14.9 &22.1 \\ 
RandN+Outer & 4.6 &91.0 & 13.9 &74.8 \\ 
FixedN+Inner & 5.6 &17.1 &15.8  &15.5 \\ 
FixedN+Outer & 4.5 &\textbf{92.0} &14.4  & 87.4\\ 
CopyN+Inner & 5.6 &10.2 & 15.3 & 16.2\\ 
CopyN+Outer & 4.8 &87.8 & 14.3 &83.1 \\ 
\hline
\end{tabular}
\end{table}
\subsection{Ablation Study}

In this section, we discuss the effect of different poisoning  methods, data selection methods and poisoning rates in our attack. All of the experiments are conducted for 10 times. 

\textbf{The effect of poisoning methods.} As shown in Figure \ref{result}, the \emph{Outer} Method has a better performance than \emph{Inner} Method both on EER and ASR. The reason of this phenomenon may come from (\ref{gongshi}). The loss function minimizes the distance between the utterances of the same speaker and maximizes the distance between the utterances of different speakers. The \emph{Inner} method substitutes the utterance, so that a smaller distance between the same speaker can bring the attacker closer to the speaker, but at the same time, a larger distance between different speakers also brings the distance between the different attacker's utterances, which limits the effect of the attack. The \emph{Outer} method, however, inserts the attacker's utterance and adds a backdoor loss externally to avoid this problem.


\textbf{The effects of data selection methods.} 
At high poisoning rate, all three selection methods can achieve a good results. However, at low poisoning rate, \emph{CopyN} and \emph{FixedN} perform better. This is because \emph{CopyN} and \emph{FixedN} only require the model to memorize one or N fixed utterances of the attacker, while \emph{RandN} requires the model to memorize arbitrary utterances of the attacker. \emph{RandN} is more difficult to attack, and thus may be difficult to get effective at low poisoning rate.


\textbf{The effects of poisoning rates.} 
With poisoning rate increasing, both the EER and ASR of the \emph{Outer} method increased while the ASR of the \emph{Inner} method is less stable. Moreover, it can be seen that at a low poisoning rate (e.g. 0.01), the \emph{Outer} method can still achieve a high attack success rate.


\section{Conclusion}
\label{sec:CONCLUSION}

In this paper, we introduce a new backdoor attack method against speaker verification named Universal Identity attack under the open-set scenario. With this method, we solve the problem in speaker verification attack that enrolled speakers do not appear in training phase, which allows attacker impersonate any legitimate user. With experiments on TIMIT and Voxceleb datasets, we have demonstrated the effectiveness and flexibility of the attack. 


\section{Acknowledgements}
This research work has been funded by the Joint Funds of the National Natural Science Foundation of China (Grant No.U21B2020).
\newpage

\bibliographystyle{IEEEtran}

\bibliography{mybib}

\end{document}